\documentclass[aps,prb,twocolumn,superscriptaddress,longbibliography,nobalancelastpage,10pt]{revtex4-2}
\usepackage{xcolor,amsthm,amsmath,amssymb,amsxtra,amsfonts,dsfont,graphicx,bm,bbm,bbold,braket,natbib,booktabs}
\usepackage[colorlinks,citecolor=blue,linkcolor=blue,urlcolor=blue]{hyperref}

\usepackage[normalem]{ulem}
\usepackage[utf8]{inputenc}
\usepackage{microtype}

\usepackage[percent]{overpic}
\usepackage{placeins}
\usepackage[caption=false]{subfig}
\usepackage[capitalize]{cleveref}
\crefname{appendix}{App.}{App.}

\newcommand*\dd{\mathop{}\!\mathrm{d}}
\DeclareMathOperator{\tr}{tr}

\DeclareMathOperator{\E}{\mathbb{E}}
\DeclareMathOperator{\diag}{diag}

\DeclareMathOperator*{\argmin}{arg\,min}

\newcommand{\phantomsubfloat}[1]{%
    {%
        \captionsetup[subfloat]{farskip=0pt,captionskip=0pt}
        \captionsetup[subfigure]{labelformat=empty}
        \subfloat{#1}
    }
}

\pdfoutput=1

\begin{document}

\title{Variational Neural and Tensor Network Approximations of Thermal States}

\author{Sirui Lu$^{1,2}$, Giacomo Giudice$^{1,2,3}$, and J. Ignacio Cirac$^{1,2}$}
\affiliation{Max-Planck-Institut f\"ur Quantenoptik, Hans-Kopfermann-Str.\ 1, D-85748 Garching, Germany}
\affiliation{Munich Center for Quantum Science and Technology (MCQST), Schellingstr. 4, D-80799 M\"unchen, Germany}
\affiliation{PlanQC GmbH, M\"unchener Str. 34, 85748 Garching, Germany}
\date{\today}

\begin{abstract}
    We introduce a variational Monte Carlo algorithm for approximating finite-temperature quantum many-body systems, based on the minimization of a modified free energy.
    This approach directly approximates the state at a fixed temperature, allowing for systematic improvement of the ansatz expressiveness without accumulating errors from iterative imaginary time evolution.
    We employ a variety of trial states---both tensor networks as well as neural networks---as variational Ans\"atze for our numerical optimization. We benchmark and compare different constructions in the above classes, both for one- and two-dimensional problems, with systems made of up to $N=100$ spins.
    Our results demonstrate that while restricted Boltzmann machines show limitations, string bond tensor network states exhibit systematic improvements with increasing bond dimensions and the number of strings.
\end{abstract}

\maketitle

\section{Introduction}
Understanding quantum many-body systems at finite temperature is a fundamental problem, with applications ranging from condensed matter to materials science.
Traditional quantum Monte Carlo methods~\cite{DerLinden1992Quantum,Sandvik2010Computational} offer efficient computation of thermal properties, but are hindered by the sign problem in low-temperature fermionic or frustrated systems~\cite{Troyer2005Computational,Henelius2000Sign}.
A sign-problem-free method is the variational method, which seeks the optimal ansatz state through an algorithm guided by the variational principle.
At zero temperature, the combination of matrix product states (MPS) and the density matrix renormalization group algorithm has proven particularly effective, establishing it as a standard for one-dimensional quantum systems~\cite{Schollwock2011Densitymatrix}.
Yet, developing variational methods that can effectively tackle finite-temperature problems in higher dimensions continues to be an open challenge.
For example, while the ground state of the two-dimensional (2D) Hubbard model at half-filling is well understood~\cite{Zheng2017Stripe,Qin2020Absence}, the low-temperature regime remains elusive.
This work proposes a robust variational Monte Carlo algorithm for approximating finite-temperature states and introduces efficient Ans\"atze, leveraging recent innovations in tensor networks and neural networks~\cite{Glasser2018NeuralNetwork, Carleo2016Solving, Melko2019Restricted}.

Tensor network states, grounded in solid theoretical foundations~\cite{Hastings2006Solving, Wolf2008Area, Molnar2015Approximating}, efficiently represent short-range equilibrium quantum many-body systems.
Matrix product states (MPS) and matrix product operators (MPO) have been successful in representing one-dimensional (1D) systems at finite temperatures~\cite{Verstraete2004Matrix, Murg2005Efficient, Feiguin2005Finitetemperature, White2009Minimally, Stoudenmire2010Minimally}.
Despite recent advances~\cite{Chen2018Exponential, Bruognolo2017Matrix, Li2019Thermal, Li2023Tangent}, applying MPS or MPO to two dimensions is computationally challenging because their inherent 1D topology assumes a different notion of locality from typical local Hamiltonians.
Their natural generalizations in 2D, projected entangled pair states and operators (PEPS and PEPO)~\cite{Verstraete2004Renormalization}, face challenging computational costs~\cite{Verstraete2006Criticality, Schuch2007Computational, Lubasch2014Algorithms, Lubasch2014Unifying}, and have mainly been used in infinite systems~\cite{Czarnik2012Projected, Czarnik2019Time, Kshetrimayum2019Tensor, Poilblanc2021Finitetemperature, Schmoll2024Finite}.
This sparked the development of extensions of tensor network wavefunctions, such as string-bond states~(SBS)~\cite{Schuch2008Simulation, Sfondrini2010Simulating, Glasser2018NeuralNetwork} and entangled plaquette states~(EPS)~\cite{Schuch2008Simulation, Mezzacapo2009Groundstate, Changlani2009Approximating, Al-Assam2011Capturing, Thibaut2019Longrange}, for which expectation values can be efficiently computed using Monte Carlo methods.
In this paper, we extend these constructions to represent mixed states, by means of purifications.\\

Neural network states, inspired by the success of deep learning, have emerged as a flexible variational ansatz for which expectation values can also be efficiently computed with Monte Carlo methods~\cite{Carleo2016Solving, Choo2018Symmetries, Choo2019Twodimensional, Sharir2020Deep, Vieijra2020Restricted, Nomura2021DiracType, Chen2024Empowering}.
Later, neural network states were connected with extensions of tensor networks~\cite{Glasser2018NeuralNetwork,Chen2018Equivalence}.
Despite their computational efficiency, the theoretical understanding of these classes of variational Ans\"atze in representing quantum states is still an active area of research compared to more established tensor networks~\cite{Deng2017Quantum, Deng2017Machine, Gao2017Efficient,Chen2018Equivalence, Levine2019Quantum, Lu2019Efficient, Huang2021Neural, Sharir2022Neural, Sun2022Entanglement}.
Neural networks have been extended to represent mixed states~\cite{Torlai2018LatentRBM, Hartmann2019NeuralNetwork, Vicentini2019Variational, Nagy2019Variational,Yoshioka2019Constructing,Vicentini2022Positivedefinite,Yuan2021Solving,Nomura2021Purifying}, where the challenges are more pronounced.
Initial studies on open quantum system dynamics with restricted Boltzmann machines (RBM)~\cite{Hartmann2019NeuralNetwork, Vicentini2019Variational, Nagy2019Variational, Vicentini2022Positivedefinite} have revealed discrepancies with exact results when dissipation strongly competes with unitary dynamics, even for small systems.
Using the graphical language of tensor networks, we generalize the connections between tensor network and neural network states to mixed states, allowing a better understanding of their relationships and enabling the construction of Ans\"atze inspired by both.

\begin{figure*}[t]
    \centering
    \phantomsubfloat{\label{fig:ansatz-graphical}}
    \phantomsubfloat{\label{fig:ansatz-mps}}
    \phantomsubfloat{\label{fig:ansatz-snakemps}}
    \phantomsubfloat{\label{fig:ansatz-peps}}
    \phantomsubfloat{\label{fig:ansatz-eps}}
    \phantomsubfloat{\label{fig:ansatz-sbs}}
    \phantomsubfloat{\label{fig:ansatz-rbm}}
    \includegraphics[width=\textwidth]{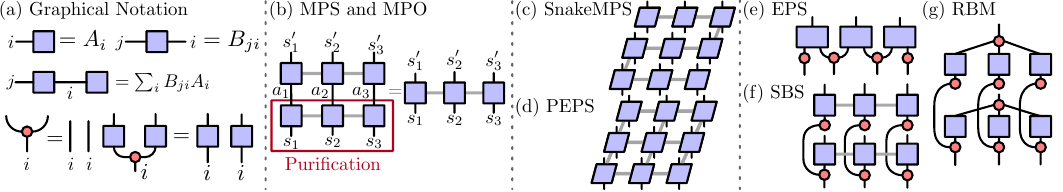}
    \vspace*{-5mm}
    \caption{\label{fig:ansatz}%
        Tensor network representation for mixed states of various tensor and neural network Ans\"atze.
        (a) Graphical notation: tensors as squares, indices as lines, contractions as line connections, and index duplication as red dots.
        (b) A matrix product operator (MPO) does not ensure the positivity of the density matrix, but a matrix product density operator (MPDO) does, via purification.
        Physical indices (downward arrows) have dimension two for spin-1/2 systems; ancillae (upward arrows) have Kraus dimension $\chi$.
        The gray lines connecting the squares have uniform bond dimensions $D$.
        All the Ans\"atze used in this work are defined and depicted by their purification.
        (c) SnakeMPS: An MPS configured in a snake-like pattern across a 2D lattice.
        (d) Projected entangled pair states (PEPS) are theoretically appealing 2D tensor networks but computationally demanding.
        (e) An entangled plaquette state (EPS) uses index duplication, where physical indices $s_i$ are shared between overlapping plaquettes, each connected to auxiliary indices $a_i$.
        (f) A string-bond state (SBS): physical indices $s_i$ are copied and reused in different MPS strings.
        (g) A restricted Boltzmann machine (RBM) state: a neural network ansatz, partially shown for the contribution from two auxiliary neurons to three spins, illustrating the duplication of both system and ancillary indices and the global purification.
    }
\end{figure*}

Current finite-temperature methods typically employ imaginary time evolution from the infinite-temperature state~\cite{White2009Minimally, Stoudenmire2010Minimally, Claes2017Finitetemperature, Chen2018Exponential, Kadow2023Isometric}, a technique well suited for tensor networks due to their theoretical guarantees and efficient truncation schemes.
However, this approach is less effective for neural networks, where evolution can only be approximated using Monte Carlo sampling~\cite{Nomura2021Purifying, Hendry2022Neural, Nys2024Realtime}, leading to potential noise and error accumulation in large or low-temperature systems.
The Gibbs free energy principle offers an alternative variational framework, but computing the von Neumann entropy is infeasible for most variational states, except for certain cases such as Gaussian states~\cite{Feynman1988Difficulties,Shi2020Variational}.
To overcome these limitations, we extend the approach of Ref.~\cite{Giudice2021Renyi} to the context of variational Monte Carlo and finite systems.
The optimization proposed is based on a modification of the free energy, such that one constructs the mixed state $\rho_{R} (\beta_R) = \arg\min_{\rho \succeq 0} F_{R}(\rho)$ where
\begin{equation}
    F_R(\rho) = \beta_R\tr(H\rho)-S_{2} (\rho).
    \label{eq:RenyiEns}
\end{equation}
Here, $H$ is the Hamiltonian, and the von Neumann entropy is replaced by the second R\'enyi entropy $S_2(\rho)=-\log \tr(\rho^2)$.
This state does not correspond to the Gibbs ensemble, but, in the thermodynamic limit, the two become locally indistinguishable at the same energy density~\cite{Giudice2021Renyi}.
Therefore, the local observables of $\rho_R$ will converge to the thermal ones as the size of the system increases.
The argument is based on the scaling of the energy variance of the two ensembles and the equivalence of ensembles~\cite{Hartmann2004Gaussian,Brandao2015Equivalence}.
We have applied this algorithm to a wide range of neural networks and tensor networks (see \cref{fig:ansatz}), and benchmarked it in 1D and 2D systems against established methods.
Our results demonstrate that the algorithm, when combined with appropriate variational Ans\"atze, has the potential to provide robust predictions for the local properties of finite-temperature quantum systems.

\section{Variational Monte Carlo}
We consider a system of $N$ spins, each of local dimension $d$, represented in the basis $\ket{\bm{s}}= \ket{s_1,\dots, s_N}$.
Given an ansatz for the mixed state $\rho=\sum_{\bm{s},\bm{s'}} \rho_{\bm{s}\bm{s'}} \ket{\bm{s}}\bra{\bm{s'}}$, we use Monte Carlo methods to compute the energy $E= \tr(H \rho)$ and the state purity $\Gamma=\tr\rho^2$.
Both are calculated as expectation values on the probability distribution defined by unnormalized diagonal matrix elements $\rho_{\bm{ss}}$,
\begin{align}
    E &= \!\!\!\!\underset{\bm{s} \sim \diag\rho}{\E}\!\left[ \sum_{\bm{s^\prime}} H_{\bm{s^\prime} \bm{s}} \frac{\rho_{\bm{s} \bm{s^\prime}}}{\rho_{\bm{s} \bm{s}}} \right],\label{eq:mean:E}\\
    \Gamma &= \!\!\!\!\underset{\substack{\bm{s} \sim \diag \rho \\ \bm{s^\prime} \sim \diag \rho}}{\E}\left[ \frac{|\rho_{\bm{s} \bm{s^\prime}}|^2}{\rho_{\bm{s} \bm{s}} \rho_{\bm{s^\prime} \bm{s^\prime}}} \right].
    \label{eq:mean:purity}
\end{align}
Here, $H_{\bm{s^\prime} \bm{s}}$ are the non-zero matrix elements of the Hamiltonian, which are polynomially many for local Hamiltonians.
The second R\'enyi entropy is estimated using $S_2(\rho)=-\log \Gamma(\rho)$.
This quantity is challenging to evaluate directly, since one must first compute $\Gamma$, which becomes exponentially small with system size.

To optimize our variational ansatz, we compute the gradients of the free energy with respect to the variational parameters $\theta$, combining the gradients of the energy and entropy as $\partial_{\theta}F_R(\rho)=\beta_R \partial_{\theta} E+ \partial_{\theta} S_2(\rho)$.
These gradients are estimated as
\begin{align}
    \partial_\theta E         &= \!\!\!\!\underset{\bm{s} \sim \diag\rho}{\E}\!\left[\sum_{\bm{s^\prime}} H_{\bm{s^\prime} \bm{s}} \frac{\partial_\theta \rho_{\bm{s} \bm{s^\prime}}}{\rho_{\bm{s} \bm{s}}} - E \frac{\partial_\theta \rho_{\bm{s} \bm{s}}}{\rho_{\bm{s} \bm{s}}}  \right],\\
    \partial_\theta S_2(\rho) &= 2 \!\!\!\!\!\!\!\!\underset{\bm{s}\bm{s^\prime} \sim \diag\rho^2}{\E}\!\left[\frac{\partial_\theta \rho_{\bm{s} \bm{s^\prime}}}{\rho_{\bm{s} \bm{s^\prime}}} \right] -  2 \!\!\!\!\underset{\bm{s} \sim \diag\rho}{\E} \!\left[\frac{\partial_\theta \rho_{\bm{s} \bm{s}}}{\rho_{\bm{s} \bm{s}}} \right].
\end{align}
Here, $E = \tr(H\rho)$ is the energy expectation value, which can be estimated using~\cref{eq:mean:E}.
The key idea here is that while estimating $S_2(\rho)$ is not scalable to large systems, gradients $\partial_{\theta} S_2(\rho)$ can be efficiently estimated by sampling not only $\rho_{\bm{ss}}$ but also the distribution proportional to $|\rho_{\bm{s}\bm{s^\prime}}|^2$.

\section{Variational Ans\"atze}
We use the graphical notation of \cref{fig:ansatz-graphical} to construct variational Ans\"atze for the density operator for which $\rho_{\bm{s}\bm{s'}}$ and its gradients $\partial_{\theta} \rho_{\bm{s}\bm{s'}}$ can be computed efficiently, up to a normalization factor.
MPO and PEPO have been shown to be efficient representations of any finite-temperature thermal states of local Hamiltonians in 1D and 2D~\cite{Hastings2006Solving, Schuch2008Peps, Wolf2008Area, Molnar2015Approximating, GuthJarkovsky2020Efficient, Alhambra2021Locally}.
For 2D problems, the computational cost of applying PEPS~[\cref{fig:ansatz-peps}] is prohibitive.
Their exact contraction scales exponentially; advanced approximation techniques scale polynomially, yet poorly for practical purposes~\cite{Verstraete2006Criticality, Schuch2007Computational, Lubasch2014Algorithms, Lubasch2014Unifying}.
In practice, the most effective solution has been adapting the MPO [\cref{fig:ansatz-mps}] on the 2D lattice in a snake pattern [\cref{fig:ansatz-snakemps}], but the bond dimension required typically grows exponentially in the length of the lattice.
To ensure that $\rho$ is Hermitian and positive semidefinite, we use a purification $\ket{\Psi}$ such that $\rho=\tr_{\mathcal{A}}(\ket{\Psi}\bra{\Psi})$ is recovered by tracing over an ancillary system $\mathcal{A}$.
This is illustrated in Fig.~1(b) for the case where $\ket{\Psi}$ is represented by an MPS, giving rise to a matrix product density operator (MPDO)~\cite{Verstraete2004Matrix}. An MPDO is graphically depicted as a ladder-like structure with bond indices of dimension $D$ and a virtual bond index of dimension $\chi$.
Notably, the MPDO is not equivalent to the mixed state represented by an MPS as the purification, as there are examples of MPDOs that have no local MPS purification of given bond dimension~\cite{Cuevas2013Purifications}. Conversely, any locally purified MPS/PEPS of bond dimension $D$ can be represented by an MPDO/PEPDO of bond dimension $D^2$. All the Ans\"atze used in this work are defined and depicted by their purification.

Purifications allow us to introduce a versatile set of variational Ans\"atze.
These Ans\"atze are constructed by dividing the lattice into $P$ overlapping subblocks, each containing $n_p$ spins indexed by $\bm{s}_p$ and associated with a unique set of ancillae $\bm{a}_p$.
The global wave function is then expressed as the product of the overlapping subblock wave functions as $\Psi(\bm{s}, \bm{a})=\prod_{p=1}^P \phi^{[p]}_{\bm{s}_p,\bm{a}_p}$, allowing different correlations of the system to be captured in a way similar to the products of so-called \emph{experts} in machine learning~\cite{Hinton2002Training}.
The density matrix of the system is obtained by tracing out all the ancillae, resulting in a factorized form 
\begin{equation}
    \rho_{\bm{ss'}} = \prod_{p=1}^P \sum_{\bm{a}_p} \phi^{[p]}_{\bm{s}_p,\bm{a}_p}(\phi^{[p]}_{\bm{s}_p,\bm{a}_p})^*.
\end{equation}
Each subblock $\phi^{[p]}_{\bm{s}_p,\bm{a}_p}$ can be represented by a full tensor, a small tensor network, or a neural network such as an RBM.
These give rise to purification Ans\"atze represented by EPS~\cite{Mezzacapo2009Groundstate}, SBS~\cite{Schuch2008Simulation}, or RBM~\cite{Torlai2018LatentRBM}.

The essential tool for representing EPS and SBS purifications as tensor networks in \cref{fig:ansatz-eps,fig:ansatz-sbs} is the copy tensor [\cref{fig:ansatz-graphical}], which duplicates indices $\ket{s_i}$ into $\ket{s_i,s_i}$.
In EPS, physical indices are copied and reused in different plaquettes, each associated with local ancillae represented by upward indices.
The overlapping nature of EPS allows them to improve upon the mean-field approximations and capture correlations between subblocks.
In SBS, physical indices are duplicated and fed into different MPS strings, so that the choice of string patterns determines the expressiveness of the ansatz.
When the underlying lattice is beyond one dimension, a natural strategy is to adapt multiple long strings that snake through the entire lattice.
We call this ansatz SnakeSBS, and it contains the SnakeMPS ansatz of \cref{fig:ansatz-snakemps} as a special case of one string.
As can be inferred for both EPS and SBS from \cref{fig:ansatz-eps,fig:ansatz-sbs}, the partial trace of the ancillae can be carried out in parallel for each plaquette or string.

An RBM purification~\cite{Torlai2018LatentRBM, Hartmann2019NeuralNetwork, Vicentini2019Variational, Nagy2019Variational} involves three sets of binary units: $s_j, j\in \{1,\ldots, N\}$, the configurations of the physical spins; hidden units $h_i, i\in \{1,\ldots,\alpha N\}$, introducing correlations between the physical spins; and ancillary units, $a_k, k\in \{1,\ldots,\beta N\}$, acting as the purification.
The relationship between RBM states and MPS, EPS, and SBS has been established in previous studies~\cite{Gao2017Efficient, Glasser2018NeuralNetwork, Chen2018Equivalence, Glasser2020Probabilistic}.
Using the purification and the graphical notation, these connections are straightforwardly generalized to their mixed-state counterparts.
Ignoring single-body terms, which can be easily absorbed in the tensor network representation, the RBM wave function is given by $\Psi(\bm{s}, \bm{a}) \propto \prod_{i=1}^{\alpha N} X_i(\bm{s}) \prod_{k=1}^{\beta N} Y_k(\bm{s}, a_k)$, where $X_i(\bm{s})=\cosh (b_i^{h} + \sum_{j} W_{ij}^h s_j )$ and $Y_k(\bm{s}, a_k)=\exp (\sum_{j} W_{kj}^a a_k s_j)$, where $b_i^{h}$ and $W_{ij}^h$ are bias and weight parameters.
The factor $X_i(\bm{s})$ entangles physical spins without involving ancillae. This function can be represented by a matrix product state $\tr \left(\prod_{i\in j} A^{s_j}_{i,j}\right)$ with diagonal matrices of bond dimension two, as proven in~\cite{Glasser2018NeuralNetwork}. 
A step by step derivation is given in \cref{app:ansatz-rbmdo}.
The factor $Y_k(\bm{s}, a_k)$ entangles the system with the ancillae.
We illustrate its representation as a tensor network in \cref{fig:ansatz-rbm} and observe that such purifications are global, with a single ancilla attached to each string, similar to the case of EPS.
The distinction lies in the fact that EPS use dense vectors in their purification, while RBM use a specific function whose physical indices are decoupled.
This is in contrast to MPS and PEPS, where ancillae and systems are treated on an equal footing.
By directly parameterizing \cref{fig:ansatz-rbm} with general tensors, the RBM can be generalized to handle larger local dimensions [More details of the derivations are given in \cref{app:ansatz-sbdo}].

\begin{figure}[t]
    \centering
    \begin{overpic}[]{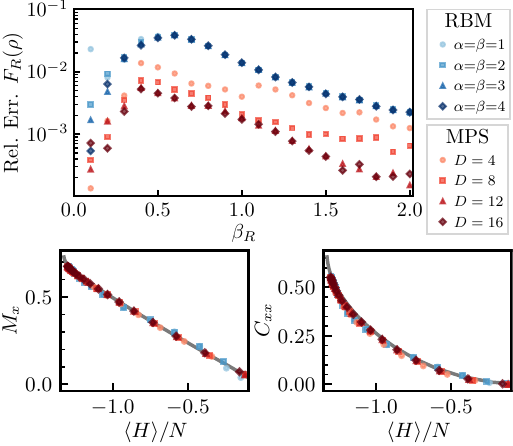}
        \put(0, 84){(a)\phantomsubfloat{\label{fig:1d-results-a}}}
        \put(5, 35){(b)\phantomsubfloat{\label{fig:1d-results-b}}}
        \put(56, 35){(c)\phantomsubfloat{\label{fig:1d-results-c}}}
    \end{overpic}
    \vspace*{-5mm}
    \caption{%
        Numerical results for approximating the R\'enyi ensemble of the non-integrable Ising model with $(J,h_z,h_x)=(1.0,0.5,1.05)$ using RBM of neural densities $\alpha=\beta=1,2,3,4$ and MPS of bond dimension $D=4,8,12,16$ and Kraus dimension $\chi=2$.
        (a) Relative error in the R\'enyi free energy as a function of $\beta_R$.
        MPS Ans\"atze converge to the exact solution as the bond dimension increases.
        RBM Ans\"atze saturate for all $\beta_R$.
        (b) Expectation value of average transverse magnetization $M_x$ and (c) two-point correlation function $C_{xx}$ as functions of energy density $\braket{H}/N$ for a system size of $N=100$.
        The results converge to the MPO results representing the Gibbs ensemble, obtained by approximating $e^{-\beta H}$ as an MPO using the TEBD algorithm to evolve from the infinite temperature state.
    }
    \label{fig:1d-results}
\end{figure}

\begin{figure*}[tb]
    \centering
    \begin{overpic}[]{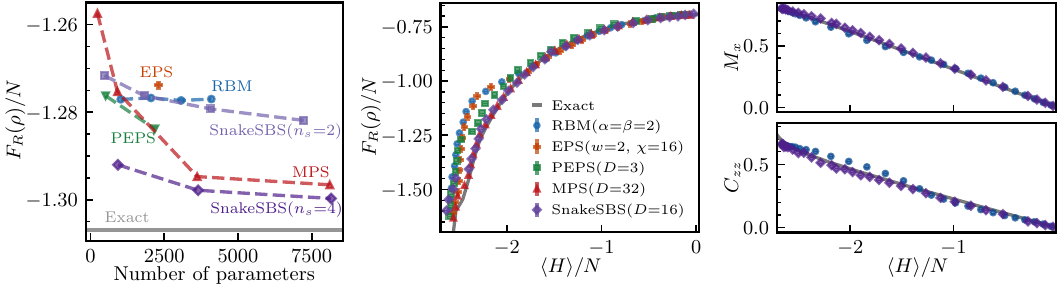}
        \put(5.25, 26.25){(a)\phantomsubfloat{\label{fig:2d-results-a}}}
        \put(38.5, 26){(b)\phantomsubfloat{\label{fig:2d-results-b}}}
        \put(70, 26){(c)\phantomsubfloat{\label{fig:2d-results-c}}}
        \put(70, 14){(d)\phantomsubfloat{\label{fig:2d-results-d}}}
    \end{overpic}
    \vspace*{-5mm}
    \caption{%
        Approximation of the R\'enyi ensemble for the 2D transverse field Ising model on a square lattice with open boundary condition at $J=-1$, $h_z=0$ using various Ans\"atze.
        (a) Variational free energy per site for $N=4\times 4$ as a function of the number of parameters for different Ans\"atze at $h_x=3.0$ and $\beta_R=0.4$.
        Most Ans\"atze improve with increasing number of parameters, except for the RBM ansatz.
        (b) Convergence to exact solutions for different Ans\"atze plotted as a function of the energy density $\braket{H}/N$ at $h_x=2.5$.
        For both (a) and (b), the solid lines correspond to the exact solution for the R\'enyi ensemble obtained by exact diagonalization.The data points are for the R\'enyi ensemble obtained by our variational approach.
        (c) Transverse magnetization $M_x$ and (d) nearest-neighbor spin-spin correlation $C_{zz}$ as functions of energy density, computed for a larger system of $N=10\times 10$ spins using SnakeSBS with $n_s=4,~D=20$ and RBM Ans\"atze at $h_x=2.5$. In (c-d), we also show QMC results for the Gibbs ensemble (gray lines) for comparison.
    }
    \label{fig:2d-results}
\end{figure*}

\section{Numerical Results in 1D}
Our variational algorithm's performance is first assessed in one-dimensional systems, using the nearest-neighbor Ising model with a Hamiltonian
\begin{equation}
    H=J\sum_{\braket{i,j}} \sigma_{i}^z\sigma_{j}^z+h_z\sum_{i=1}^{N} \sigma_{i}^z-h_x\sum_{i=1}^{N} \sigma_{i}^x.
    \label{eq:HamIsing}
\end{equation}
We benchmark our algorithm at a non-integrable point $J=1,\ h_z=0.5,\ h_x=1.05$ with an open boundary condition~\footnote{Additional numerical results are shown in \cref{app:additional-numerical-results}.}.
We first examine a system of $N=16$ spins, comparing with exact diagonalization results.
Although restricted to small systems, the possibility of estimating the free energy [\cref{eq:mean:E,eq:mean:purity}] provides a way to benchmark different Ans\"atze, which is generally prohibited for the algorithm based on imaginary time evolution.
We test both RBM and MPS, with results shown in~\cref{fig:1d-results-a}.
Increasing the bond dimension of the MPS ansatz improves convergence to the exact solution, while surprisingly, this does not happen for RBM.
The universal approximation properties of RBM~\cite{Montufar2010Refinements} might lead one to anticipate similar capabilities for their purified states, and, by extension, for the reduced density matrices.
However, to enable analytical partial trace, the purification units of this RBM ansatz are integrated within the hidden layer rather than within the visible layer.
This violates the conditions demanded by the universal approximation theorem~\cite{Montufar2010Refinements}.
The universal approximation theorem holds if the density operator is modeled using a deep Boltzmann machine (DBM)~\cite{Nomura2021Purifying}.
But the optimization of DBM is computationally expensive~\cite{Gao2017Efficient,Carleo2018Constructing}, which restricts the attainable system sizes~\cite{Nomura2021Purifying,Carleo2018Constructing}.

To evaluate the scalability of the ansatz and the algorithm, we scale up to $N=100$ spins in \cref{fig:1d-results-b,fig:1d-results-c}.
Because our algorithm is variational, using a few thousand samples is sufficient for computing the gradients, an order of magnitude lower than what is required to accurately simulate the time evolution reported in previous works~\cite{Vicentini2022Positivedefinite,Nys2024Realtime}.
Two local observables, the transverse magnetization $M_x=\sum_i\braket{\sigma^x_i}/N$ and the two-point correlation function $C_{xx}=\sum_{i}\braket{\sigma^x_i \sigma^x_{i+1}}/N$, are measured on optimized RBM and MPS Ans\"atze.
To compare with the results of the Gibbs ensemble, they are plotted against the energy density $\braket{H}/N$.
As shown in \cref{fig:1d-results-b,fig:1d-results-c}, both results closely match the MPO results.

\section{Numerical results in 2D}
We turn to the two-dimensional transverse field Ising model in \cref{eq:HamIsing}, with parameters
$J=1,h_z=0,h_x=3$ that are close to the critical value $|h_x^c|\approx 3.044$~\cite{Blote2002Cluster}.
We first benchmark the RBM, MPS, PEPS, and our newly introduced EPS and SnakeSBS Ans\"atze for $N=16$ spins arranged on a $4 \times 4$ square lattice with open boundary conditions.
These Ans\"atze are first evaluated in \cref{fig:2d-results-a} at an intermediate temperature $\beta_R=0.4$, which is particularly challenging for variational methods due to strong competition between quantum and thermal fluctuations.
The RBM show a saturation effect with a relative error of approximately $2.2\%$, suggesting similar representational limits.
The EPS ansatz is found to be less accurate, with a relative error of $2.5\%$, while being difficult to scale up because the number of parameters scales exponentially with the size of the plaquettes.
The PEPS ansatz, expected to accurately capture thermal physics, approaches the exact solution at bond dimensions $D=2$ and $D=3$ with relative errors of $2.3\%$ and $1.7\%$.
However, its scalability is challenged by the expensive contraction procedure.
We find that the MPS ansatz improves consistently with increasing bond dimensions, with relative errors of $3.7\%$, $2.4\%$, $0.9\%$, and $0.4\%$ for $D=4,~8, 16, 32$, respectively.
The SnakeSBS ansatz uses multiple MPS strings that snake through the lattice horizontally and vertically [see Fig.~4 for an illustration of the string configurations]. We find improvements with the SnakeSBS ansatz in $F_R(\rho)$ with increasing bond dimensions $D$ and number of strings $n_s$, with relative errors decreasing from $2.6\%$ to $1.9\%$ for $n_s=2$ strings and from $1.1\%$ to $0.5\%$ for $n_s=4$ strings when increasing from $D=4$ to $D=12$.

Next, we set $h_x=2.5$ and plot $F_R(\rho)$ against the energy density of the state in \cref{fig:2d-results-b} to compare throughout the energy spectrum.
For each ansatz, we report only the top results from our explored parameter range.
In the higher-energy regime, most of the Ans\"atze closely match the exact results.
However, at lower energies, discrepancies emerge, especially for RBM and EPS.
In this regime, the MPS and SnakeSBS Ans\"atze show better convergence to the exact results with higher bond dimensions.
Although MPS attain lower $F_R(\rho)$ with a larger bond dimension $D=32$, we have found that SnakeSBS are the most scalable Ans\"atze.
We optimize the SnakeSBS and RBM Ans\"atze for a range of energy densities on a larger system of $N=100$ spins and compare the results with quantum Monte Carlo (QMC).
As shown in \cref{fig:2d-results-c,fig:2d-results-d}, the approximation for the transverse magnetization $M_x$ is found to be more accurate than that for the two-point correlation function $C_{zz}=\sum_{\langle i,j\rangle} \braket{\sigma^z_i \sigma^z_{j}}/N$. For the $10\times10$ system, the SnakeSBS ansatz with $D=20$ shows decent agreement with quantum Monte Carlo results for both the transverse magnetization $M_x$ and the nearest-neighbor correlation $C_{zz}$. The relative error in these observables is less than $2\%$ for all energy densities studied, demonstrating the ability of our method to approximate thermal state properties even for larger 2D systems.
While the Ising model we studied can be effectively addressed with QMC, variational approaches offer additional advantages, such as being capable of measuring more complex observables and not being limited by the sign problem.

\section{Conclusion and Outlook}
We proposed a variational Monte Carlo algorithm to approximate thermal states and we use it to optimize variational ansatze based on both tensor networks and neural networks.
Our approach directly approximates the state at a fixed temperature and presents a robust alternative to the imaginary time evolution method.
This direct approximation permits a controlled and gradual enhancement in the expressiveness of the ansatz to refine the accuracy of the approximation.
We tested this combination on one- and two-dimensional quantum Ising models, comparing the results with established methods.
In both 1D and 2D, we observed that RBM yield saturating results and do not converge to the exact solution.
In contrast, MPS provided perfect results in 1D.
In 2D, we found that SnakeSBS improves with increasing number of strings and bond dimensions, outperforming RBM in our comparison to the exact results.
For $N=100$ spins, the results of SnakeSBS show good agreement with those of the Gibbs ensemble, particularly for local observables, while RBM show limitations in capturing the thermal state properties accurately.

Using tensor network diagrams, we introduced a series of string-bond states with purifications.
These SBS interpolate between MPS and RBM, and can handle systems with larger local Hilbert spaces with minor modifications.
They are suitable for addressing the dynamics of open quantum systems~\cite{Verstraete2004Matrix,Hartmann2019NeuralNetwork, Vicentini2019Variational, Nagy2019Variational,Yuan2021Solving}.
Future improvements may come from using tensor network diagrams as a tool to design deeper Ans\"atze that can be contracted with Monte Carlo methods~\cite{Levine2019Quantum, Glasser2020Probabilistic, Vicentini2022Positivedefinite}.
We anticipate that more expressive Ans\"atze will enable our algorithm to tackle complex problems at the forefront of experiments and theory, including three-dimensional models~\cite{Sfondrini2010Simulating}, long-range models~\cite{Deng2017Quantum, Ebadi2021Quantum, Chen2023Continuous}, frustrated models~\cite{Sfondrini2010Simulating, Mezzacapo2010Groundstate}, and chiral models~\cite{Nielsen2013Local, Glasser2018NeuralNetwork, Huang2021Neural}, whose ground-state properties have been promisingly investigated with neural network or string-bond states. Extending our approach to fermionic systems like the Hubbard model is feasible using either a Jordan-Wigner transformation or fermionic MPS.

\begin{acknowledgments}
    We would like to thank P.~Emonts for addressing all inquiries related to Monte Carlo and supplying additional Quantum Monte Carlo data.
    We also acknowledge illuminating discussion with T.~Shi, I.~Glasser, M.-C.~Ba\~{n}uls and F.~Vicentini.
    Our implementation uses the codebase of Jax~\cite{Jax}, Flax~\cite{Flax}, and NetKet 3~\cite{Vicentini2022NetKet}.
    Quantum Monte Carlo and tensor network benchmarks were performed with ITensor.jl~\cite{Fishman2022ITensor} and ALPS~\cite{Bauer2011ALPS}.
    This research is part of the Munich Quantum Valley, which is supported by the Bavarian state government with funds from the Hightech Agenda Bayern Plus.
    The work is partially supported by the Deutsche Forschungsgemeinschaft (DFG, German Research Foundation) under Germany's Excellence Strategy - EXC-2111 - 390814868.
\end{acknowledgments}

\FloatBarrier

\appendix



\section{Applying the R\'enyi Free Energy Principle with Variational Monte Carlo}
\label{app:vmc}

\subsection{Estimation and variational minimization using Monte Carlo}

To find the thermal state $\rho_G(\beta)$ at a given inverse temperature $\beta$, it is in principle possible to minimize the Gibbs free energy
\begin{equation}
    F_G(\rho)=\beta \tr(H\rho)-S(\rho),
\end{equation}
where $S(\rho)=-\tr(\rho\log\rho)$ denotes the von Neumann entropy.
However, the computation of $S(\rho)$ is demanding as it requires diagonalizing the many-body density matrix $\rho$.
An alternative, which is computationally more tractable, utilizes the $\alpha$-R\'enyi entropy, $S_{\alpha}(\rho)=\frac{1}{1-\alpha} \log \tr \rho^{\alpha}$, leading to the definition of the $\alpha$-R\'enyi ensemble through a modified free energy $F_\alpha$~\cite{Giudice2021Renyi}
\begin{align}
    F_\alpha(\rho)       &= \beta_\alpha \tr(H\rho)-S_{\alpha}(\rho),\\
    \rho_{\alpha}(\beta) &= \argmin_{\rho \succeq 0, \tr \rho=1} F_{\alpha}(\rho).
\end{align}
This variational approach requires minimization over density operators that are (i) positive semidefinite, (ii) Hermitian, and (iii) normalized.
We focus on the case $\alpha=2$, known as the second R\'enyi ensemble.
Given that our variational Ans\"atze for $\rho_{\bm{ss'}}$ are defined up to a normalization factor, we incorporate this factor into the R\'enyi free energy expression $F_R(\rho)$ for $\alpha=2$
\begin{equation}
    \begin{aligned}
        F_R(\rho) & = \beta_R \braket{H}_\rho - S_2(\rho)                                                      \\
                  & = \beta_R \frac{\tr(H \rho)}{\tr{\rho}} + \log \frac{\tr(\rho^2)}{\left(\tr\rho\right)^2}.
        \label{eq:F2}
    \end{aligned}
\end{equation}
Variational Monte Carlo works by representing quantities of interest as expectation values over a specific probability distribution.
These values are then approximated as statistical averages of samples, which are obtained using Markov chain Monte Carlo methods.

The first term of \cref{eq:F2}, the energy, can be expressed and estimated as follows:
\begin{equation}
    \begin{aligned}
        \braket{H} & = \frac{\sum_{\bm{s}, \bm{s'}} H_{\bm{s' s} } \rho_{\bm{ss'}}}{\sum_{\bm{s}} \rho_{\bm{ss}}}
        = \frac{\sum_{\bm{s}, \bm{s'}} \rho_{\bm{ss}} H_{\bm{s' s} } \frac{\rho_{\bm{ss'}}}{\rho_{\bm{ss}}}}{\sum_{\bm{s}} \rho_{\bm{ss}}} \\
                   & = \E_{\bm{s} \sim \diag\rho} \left[ \sum_{\bm{s'}} H_{\bm{s' s} } \frac{\rho_{\bm{ss'}}}{\rho_{\bm{ss}}} \right]      \\
                   & = \E_{\bm{s} \sim \diag\rho}\left[ E_{\mathrm{loc}} (\bm{s})\right].
    \end{aligned}
    \label{eq:energy_estimator}
\end{equation}
Here, the local energy is defined as
\begin{equation}
    E_{\mathrm{loc}}(\bm{s})=\frac{\bra{\bm{s}}\rho H\ket{\bm{s}}}{\bra{\bm{s}}\rho\ket{\bm{s}}}=\sum_{\bm{s}^{\prime}} H_{\bm{s' s}} \frac{\rho_{\bm{ss'}}}{\rho_{\bm{ss}}}.
    \label{eq:local_estimator}
\end{equation}
\cref{eq:energy_estimator,eq:local_estimator} are not only applicable for estimating the energy, but can also be used for other Hermitian observables, such as magnetization and correlation functions.
For a Hamiltonian $H$ (or any other arbitrary observable) that is local or a sum of local terms, the local estimators of \cref{eq:local_estimator} involve only a polynomial number of nonzero terms, thus enabling efficient evaluation.
It is worth noting that the Metropolis-Hastings algorithm, which only requires computing the ratio of the probability densities between the current and proposed states, can be used to sample from unnormalized probability densities, such as the unnormalized $\rho_{\bm{ss}}$ and $|\rho_{\bm{ss'}}|^2$.

The second term of \cref{eq:F2}, the second R\'enyi entropy, requires computing the purity $\Gamma = \tr \rho^2/(\tr \rho)^2$ before taking the logarithm.
The purity can be estimated as
\begin{equation}
    \begin{aligned}
        \Gamma & = \frac{\sum_{\bm{s}, \bm{s'}} \rho_{\bm{s' s} } \rho_{\bm{ss'}}}{\left( \sum_{\bm{s}} \rho_{\bm{ss}} \right) \left( \sum_{\bm{s'}} \rho_{\bm{s' s'}} \right)}                                                                           \\
               & = \frac{\sum_{\bm{s}, \bm{s'}} \rho_{\bm{ss}} \rho_{\bm{s' s'}} \frac{\rho_{\bm{s' s} } \rho_{\bm{ss'}}}{\rho_{\bm{ss}} \rho_{\bm{s' s'}}}}{\left( \sum_{\bm{s}} \rho_{\bm{ss}} \right) \left( \sum_{\bm{s'}} \rho_{\bm{s' s'}} \right)} \\
               & = \E_{\substack{\bm{s} \sim \diag\rho                                                                                                                                                                                                    \\ \bm{s'} \sim \diag\rho}} \left[ \frac{|\rho_{\bm{ss'}}|^2}{\rho_{\bm{ss}} \rho_{\bm{s' s'}}} \right].
    \end{aligned}
\end{equation}
The diagonal elements of $\rho$ can be sampled independently.
This equation is equivalent to employing the so-called swap trick to compute the 2-R\'enyi entropy as $\Gamma = \braket{\text{SWAP}}_{\rho \otimes \rho}$ on two copies of $\rho$.

The gradient of the cost function, necessary for the variational optimization, can be expressed as well as an expectation value suitable for Monte Carlo sampling.
Assuming that the density operator is parametrized by a set of parameters $\bf{\theta}$, we compute the gradients with respect to these parameters.
The gradient of the energy is given by
\begin{equation}
    \begin{aligned}
          & \partial_{\theta} \braket{H} = \tr\left[\left(\frac{H - \braket{H}}{\tr \rho}  \right) \partial_{\theta}\rho\right]                                                                                                                          \\
        = & \E_{\bm{s} \sim \diag\rho} \left[\sum_{\bm{s'}} H_{\bm{s' s} } \frac{\partial_{\theta}\rho_{\bm{ss'}}}{\rho_{\bm{ss}}} \right] - \braket{H} \E_{\bm{s} \sim \diag\rho} \left[\frac{\partial_{\theta}\rho_{\bm{ss}}}{\rho_{\bm{ss}}} \right].
    \end{aligned}
    \label{eq:energy_gradient}
\end{equation}
The gradient of the purity is given by
\begin{equation}
    \partial_{\theta}\Gamma = \tr \left[\left(\frac{2 \rho}{\left(\tr\rho\right)^2}
        - \frac{2 \Gamma}{\tr \rho}\right) \partial_{\theta}\rho \right].
\end{equation}
Using this, we can calculate the gradients of the logarithm purity
\begin{equation}
    \begin{aligned}
        \partial_{\theta}\log{\Gamma} & =  \frac{\partial_{\theta}\Gamma}{\Gamma} = \tr \left[\left(\frac{2 \rho}{\Gamma\left(\tr\rho\right)^2}
        - \frac{2}{\tr \rho}\right) \partial_{\theta}\rho \right]                                                                                                                                                                                                \\
                                      & = \tr \left[\left(\frac{2 \rho}{\tr\rho^2}
        - \frac{2}{\tr \rho}\right) \partial_{\theta}\rho \right]                                                                                                                                                                                                \\
                                      & = 2 \E_{\bm{s}\bm{s^\prime} \sim \diag\rho^2}\left[\frac{\partial_{\theta}\rho_{\bm{ss'}}}{\rho_{\bm{ss'}}} \right] -  2 \E_{\bm{s} \sim \diag\rho}\left[\frac{\partial_{\theta}\rho_{\bm{ss}}}{\rho_{\bm{ss}}} \right].
    \end{aligned}
    \label{eq:log_purity_gradient}
\end{equation}
Finally, the gradient of the functional $F_R(\rho)$ with respect to the variational parameters $\theta$ is expressed as
\begin{equation}
    \partial_{\theta}F_R(\rho)=\beta_R \partial_\theta \braket{H}+ \partial_\theta \log \Gamma,
    \label{eq:gradient}
\end{equation}
where $\partial_\theta \braket{H}$ and $\partial_\theta \log \Gamma$ are given by \cref{eq:energy_gradient} and \cref{eq:log_purity_gradient}, both of which are amenable to Monte Carlo estimations.
To optimize the variational parameters, we employ a gradient descent method, updating the parameters iteratively as follows:
\begin{equation}
    {\theta}^{t+1}={\theta}^{t}-\eta_t\partial_{\theta}F_R(\rho),
    \label{eq:GD}
\end{equation}
where $\eta_t$ denotes the learning rate at iteration $t$.

\subsection{Imaginary time evolution, Monte Carlo flow equation, and stochastic reconfiguration}
\label{app:FlowEqorSRmixed}

The Gibbs state can be obtained through imaginary time evolution using the equation of motion
\begin{equation}
    \frac{\dd \rho}{\dd \tau}=-\{H-\braket{H}, \rho\}.
    \label{eq:ITVMmixed}
\end{equation}
For a specific inverse temperature $\beta$, the evolution starts from the maximally mixed state $\mathbbm{1}$ and ends at time $\tau_\beta=\beta/2$
\begin{equation}
    e^{-\beta H}=e^{-\beta H / 2} \mathbbm{1} e^{-\beta H / 2}.
\end{equation}
The maximally mixed state is typically represented with a straightforward neural or tensor network structure.
Starting from these variational representations, we can approximate the imaginary time evolution by projecting \cref{eq:ITVMmixed} onto the variational manifold defined by the Ans\"atze.

The method of obtaining the Gibbs state, as described, has its limitations.
First, projection onto the variational manifold can introduce errors that accumulate over time, leading to a divergence between the numerical solution and the actual imaginary time evolution.
Second, the targeted Gibbs state is not a fixed point of the imaginary time evolution.
Consequently, any errors arising from statistical samplings or insufficient convergence of the Markov chain cannot be corrected back.
This contrasts with the ground state case, where the imaginary time evolution
\begin{equation}
    \partial_\tau \ket{\Psi(\tau)} = -(H-\braket{H}) \ket{\Psi(\tau)}
    \label{eq:ITV}
\end{equation}
has the ground state as its fixed point, $\lim_{\tau\to\infty}\ket{\Psi(\tau)} \to \ket{\Phi_0}$, and monotonically decreases the energy.
This evolution process corrects errors introduced by imperfect sampling or projection onto the variational manifold, guiding the state toward the best approximation of the ground state within the variational manifold.

To enhance the stability of the variational approach for thermal states, we introduce a nonlinear flow equation that shares the fixed-point characteristics of the imaginary time evolution for the ground state~\cite{Shi2020Variational,Giudice2021Renyi}
\begin{equation}
    \frac{\partial \rho_\tau}{\partial \tau}=-\frac{1}{2}\left\{{F}_\tau-\braket{{{F}_\tau}}, \rho_\tau \right\}.
    \label{eq:flowEq}
\end{equation}
In this equation, the R\'enyi free energy operator is ${F}_\tau =\beta_R H+\frac{2}{\tr \rho_\tau^2} \rho_\tau$, and $\braket{{{F}_\tau}}=\tr[\rho F_\tau(\rho)]$.
This evolution, as shown in \cref{eq:flowEq}, maintains the trace and positivity of the density matrix.
Similarly to imaginary evolution [\cref{eq:ITV}], the free energy of $\rho_\tau$ decreases monotonically with $\tau$.
Thus, by selecting a suitable initial density operator $\rho$ and integrating \cref{eq:flowEq} over an adequate interval, we can obtain a variational approximation of the R\'enyi ensemble as defined in \cref{eq:F2}.

To determine the equation of motion for the variational parameters, we first expand the evolved state according \cref{eq:flowEq} to first order,
\begin{equation}
    \rho_{\tau+\delta \tau} \approx e^{-\frac{\delta \tau}{2}\left(\mathcal{F}_\tau-\braket{\mathcal{F}_\tau}\right)} \rho_\tau e^{-\frac{\delta \tau}{2}\left(\mathcal{F}_\tau-\braket{ \mathcal{F}_\tau}\right)},
    \label{eq:IntFlow}
\end{equation}
and then apply the time-dependent variational principle~\cite{Haegeman2011TimeDependent}.
The ideal metric for this space of density operators is the Bures distance with the $L^1$ norm, but due to computational complexity, we opt for the $L^2$ norm, which is more manageable for Monte Carlo methods~\cite{Hartmann2019NeuralNetwork, Vicentini2019Variational, Nagy2019Variational, Nomura2021Purifying, Vicentini2022Positivedefinite}.

This approach leads to an update rule similar to the stochastic reconfiguration method~\cite{Becca2017Quantum}, where we adjust the gradient updates  of \cref{eq:GD} using the Gram matrix of the density matrix,
\begin{equation}
    \theta^{t+1}=\theta^{t}-\eta_t {G}^{-1}(\theta) \cdot \nabla_{\theta} F_R(\rho).
    \label{eq:SRGD}
\end{equation}
The Gram matrix is estimated as follows:
\begin{equation}
    \begin{split}
        G_{ij}(\theta) = &\E_{\substack{\bm{s} \sim \diag\rho\\ \bm{s'} \sim \diag\rho}}[{\Delta_{\theta_i}^*(\bm{s},\bm{s'})\, \Delta_{\theta_j}(\bm{s},\bm{s'})}] \\
                         &-\E_{\substack{\bm{s} \sim \diag\rho\\ \bm{s'} \sim \diag\rho}}[{\Delta_{\theta_i}^*(\bm{s},\bm{s'})}]\cdot \E_{\substack{\bm{s} \sim \diag\rho \\ \bm{s'} \sim \diag\rho}}[{\Delta_{\theta_j}(\bm{s},\bm{s'})}],
        \label{eq:Sijmixed}
    \end{split}
\end{equation}
where we define the logarithm derivative of the density matrix elements as
\begin{equation}
    \Delta_{\theta}(\bm{s},\bm{s'}) =\frac{\partial \log \rho_{\bm{ss'}}}{\partial \theta}= \frac{1}{\rho_{\bm{ss'}}} \frac{\partial \rho_{\bm{ss'}}}{\partial \theta}.
\end{equation}
Employing this method in our numerical optimizations has proven to reduce the required sample size for estimating the gradient and Gram matrix, allowing larger step sizes and thus expediting convergence.

In stochastic reconfiguration, a Gram matrix $G$ is used to transform the steepest descent directions in the Euclidean parameter space into the steepest descent directions in the variational state space.
This method, equivalent to the natural gradient descent method in machine learning, has been widely used in various variational Monte Carlo methods.

\section{Variational Ans\"atze}
\label{app:ansatz}

We describe in more detail the Ans\"atze that we used: matrix product states~(MPS), entangled plaquette states~(EPS), string-bond density states~(SBS), and restricted Boltzmann machine states~(RBM).
For these states defined by their purifications, we discuss how to efficiently compute the elements of the density matrix $\rho_{\bm{ss'}}$.

The area law of entanglement entropy and tensor networks go hand in hand, providing the key to the efficient representation of the ground state of local gapped Hamiltonians~\cite{Schuch2008Entropy}.
A similar story can be told for finite-temperature states with mutual information, defined as $I(A:B) = S(\rho_A)+S(\rho_B)-S(\rho_{AB})$~\cite{Wolf2008Area}.
The mutual information $I(A:B)$ measures the correlation between subsystems $A$ and $B$.
The mutual information of a pure joint state equals twice the entanglement entropy.
Therefore, if the purification satisfies the area law of entanglement, the reduced mixed state satisfies the area law for mutual information.
Gibbs states with local interactions and locally purified MPS and PEPS with constant bond dimensions both adhere to this area law of mutual information.
This suggests that successful Ans\"atze for finite-temperature states should also comply with this law.
Using tensor network diagrams, we can easily diagnose the scaling of mutual information for the Ans\"atze.
For the Ans\"atze that we use, we also discuss their scaling with respect to the mutual information.

\subsection{Matrix product states}
\label{app:ansatz-mps}

A matrix product state (MPS), represented with local ancillae for purification, is defined as
\begin{equation}
    \Psi_\text{MPS}({\bm{s},\bm{a}}) = \tr \left( \prod_{j=1}^N A^{[j]}_{s_j,a_j} \right).
\end{equation}
This can be visualized, in the case of five spins, as
\begin{equation}
    \ket{\Psi_\text{MPS}} = \vcenter{\hbox{\includegraphics[]{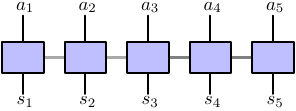}}}\ .
    \label{appfig:MPDO_puri}
\end{equation}
Tracing out the ancillae transforms the purified MPS into a mixed state, or MPDO:
\begin{equation}
    \rho_\text{MPDO}({\bm{s},\bm{s'}})=\tr \left(M^{[1]}_{s_{1},s_{1}'} \cdots M^{[N]}_{s_{N},s_{N}'} \right).
    \label{eq:MPDO}
\end{equation}
Here, local purifications imply that $M^{[j]}_{s,s'}$ are matrices of size $D^2\otimes D^2$, expressed as
\begin{equation}
    M^{[j]}_{s,s'}=\sum_{a=1}^{\chi} A_{s,a}^{[j]}\otimes (A_{s,a}^{[j]})^*,
    \label{eq:puri_MPDO}
\end{equation}
where $\chi$ is at most $D^2$.
The matrices $A_{s,a}^{[i],[j]}$ are also of size $D^2$.
An MPDO is graphically depicted as a ladder-like structure with bond indices of dimension $D$ and a virtual bond index of dimension $\chi$
\begin{equation}
    \rho_\text{MPDO} = \vcenter{\hbox{\includegraphics[]{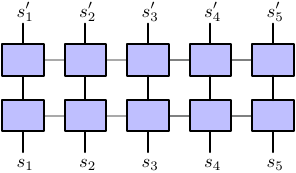}}}\ .
    \label{appfig:MPDO}
\end{equation}
Typically, $\chi$ is chosen to be comparable to the physical dimension, $\chi \sim d$.

To compute the elements of the density matrix $\rho_{\bm{ss'}}$, we utilize the chain-like structure of MPO and MPS.
We set the physical indices to specific states $\bm{s}$ and $\bm{s'}$ and decompose the matrix product density operator (MPDO) into a series of matrices.
These matrices are then multiplied in sequence, a process akin to finding the inner product in MPS, resulting in the scalar $\rho_{\bm{ss'}}$.
The computational cost of this method scales as $O(D^3)$.
This is analogous to recurrent neural networks (RNNs)~\cite{LSTM}, where data are processed sequentially.
Our implementation benefits from similar advancements in parallel computing on modern hardware, enhancing the efficiency of our MPS/MPDO calculations.

\subsection{Entangled plaquette states}
\label{app:ansatz-eps}

An entangled plaquette state~(EPS) can represent a variety of quantum states, including those relevant in quantum information theory, such as the toric code and graph states~\cite{Schuch2008Simulation}.
This has motivated the use of EPS as a variational Ansatz for quantum many-body problems.
To construct a purification, we associate an ancillary spin $a_p$ on each plaquette of the original system.
The dimension of this purifying spin is $\chi_a$, which is a variable parameter in our model.
The purified state $\ket{\Psi}$ is then expressed as a product of local terms, each corresponding to a plaquette and its associated purifying spin,
\begin{equation}
    \Psi_\text{EPS}(s_1,\ldots,s_N,a_1,\ldots,a_P) = \prod_{p=1}^P \phi^{[p]}_{\bm{s}_p,{a}_p}.
    \label{eq:EPS_puri}
\end{equation}
Here, $\phi^{[p]}_{\bm{s}_p,{a}_p}$ is the coefficient assigned to the configuration $\ket{\bm{s}_p,{a}_p}$, where $\bm{s}_p$ represents the set of physical spins in the $p$-th plaquette.
The graphical representation of this purification is depicted as follows,
\begin{equation}
    \ket{\Psi_\text{EPS}}=\vcenter{\hbox{\includegraphics[width=0.6\linewidth]{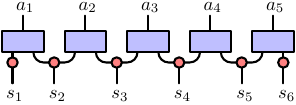}}}\label{appfig:EPS_puri},
\end{equation}
where the tensors are connected by bond indices, and each tensor has an additional purifying bond index.

We design EPS to have local purifying ancillae for each plaquette.
This enables efficient and parallel partial-trace operations.
The entangled plaquette density operator (EPDO) is obtained from the entangled plaquette state (EPS) by tracing out the auxiliary degrees of freedom as
\begin{equation}
    \begin{aligned}
          & \rho_\text{EPDO}({\bm{s},\bm{s'}}) = \sum_{\bm{a}} \prod_{p=1}^P \phi^{[p]}_{\bm{s}_p,{a}_p}\left(\phi^{[p]}_{\bm{s}_p,{a}_p}\right)^*                                                                         \\
        = & \left(\!\sum_{{a}_1} \phi^{[1]}_{\bm{s}_1,{a}_1}\left(\phi^{[1]}_{\bm{s}_1,{a}_1}\right)^*\!\right) \ldots \left(\!\sum_{{a}_P} \phi^{[P]}_{\bm{s}_P,{a}_P}\left(\phi^{[P]}_{\bm{s}_P,{a}_P}\right)^*\!\right) \\
        = & \prod_{p=1}^P\rho_{\bm{s},\bm{s}'}^{[p]}.
    \end{aligned}
\end{equation}
This simplifies to a product of plaquette density matrices for the configurations $\bm{s}_p$ and $\bm{s}_p^\prime$.
This corresponds to
\begin{equation}
    \rho_\text{EPDO}=
    \vcenter{\hbox{\includegraphics[width=0.6\linewidth]{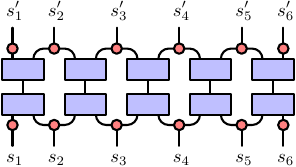}}}\label{appfig:EPDO_supp}.
\end{equation}
In this graphical representation, the physical indices are on two opposite plaquettes.
One is on the bra side and the other on the ket side.
The inner indices that connect them illustrate the partial trace.
They have a size given by the Kraus purification parameter $\chi_a$.
We show this with a plaquette size of two for simplicity.

To ensure that the density operator is of full rank, we require $(\chi_a)^{N_{p}}\geq d^N$, where $N_p$ denotes the number of plaquettes and $d$ represents the physical dimension.
This condition ensures that the purification space is sufficiently large to encompass the entire state space of the system.
Moreover, EPDO allows for weight-sharing across plaquettes, akin to the functionality in convolutional neural networks, making it suitable for systems with periodic boundary conditions.

EPDO have several notable properties that are derived from their purifications.
First, EPS are a subset of MPS~\cite{Schuch2008Simulation,Changlani2009Approximating}.
As such, EPDO also form a subclass of MPDO in one dimension.
Second, since EPS obey the area law of entanglement entropy, EPDO obey the area law of mutual information~\cite{Wolf2008Area}.

\subsection{String-bond states}
\label{app:ansatz-sbdo}

String-bond states (SBS) are tensor-network states with coefficients derived from the product of matrix product state coefficients along lattice strings.
When we include ancillae, we express the string-bond state for the purified state as
\begin{equation}
    \Psi_\text{SBS}({\bm{s},\bm{a}}) = \prod_{i=1}^{n_S} \tr \left( \prod_{j\in i}A^{[i],[j]}_{s_j,a_j} \right).
\end{equation}
In this expression, each string $i$ is a sequence selected from the set of variables $\bm{s}$, with each string linked to a set of auxiliary variables $\bm{a}$.
The tensors $A^{[i],[j]}_{s_j,a_j}$ define the amplitudes for the states of the system combined with the ancillae.
The choice of the strings and the ancillae' dimension $\chi_a$ determines the model's descriptive capability.

We derive the string-bond density operator (SBDO) by tracing out the auxiliary degrees of freedom $\bm{a}$, a process that can be performed individually for each string.
This approach resembles the method used for locally purified density operators, allowing us to consider SBDO as a type of overlapping MPDO.
For an SBDO composed of $n_S$ strings, and assuming a uniform bond dimension $D$, the density matrix elements are given by
\begin{equation}
    \begin{aligned}
        \rho_{\bm{ss'}} & =\prod_{i=1}^{n_S}\rho_{\bm{s},\bm{s}'}^{[i]}                                                                                                              \\
                        & =\prod_{i=1}^{n_S}  \tr\left(M^{[i],[j_1]}_{s_{j_1},s_{j_1}'} M^{[i],[j_2]}_{s_{j_2},s_{j_2}'} \cdots M^{[i],[j_{n_i}]}_{s_{j_{n_i}},s_{j_{n_i}}'}\right).
    \end{aligned}
    \label{eq:SBDO}
\end{equation}
In this context, local purifications imply that the matrices $M^{[i],[j]}_{s,s'}$ are of dimension $D^2\otimes D^2$ and can be expressed as
\begin{equation}
    M^{[i],[j]}_{s,s'}=\sum_{a=1}^{\chi_a} A_{s,a}^{[i],[j]}\otimes (A_{s,a}^{[i],[j]})^*.
    \label{eq:puri_SBDO}
\end{equation}
Here, $\chi_a$ is limited to $D^2$, and the matrices $A_{s,a}^{[i],[j]}$ are of size $D^2$.
Visually, for a system with $N=5$ and $n_s=2$ strings, the SBDO density matrix is represented as
\begin{equation}
    \rho_\text{SBDO}=
    \vcenter{\hbox{\includegraphics[width=0.4\linewidth]{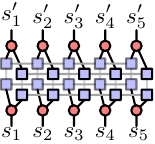}}}\label{appfig:SBDO}\ .
\end{equation}
The purification process ensures that the resulting density matrix is positive semidefinite and Hermitian by construction.

\begin{figure}[ht]
    \centering
    \begin{overpic}[]{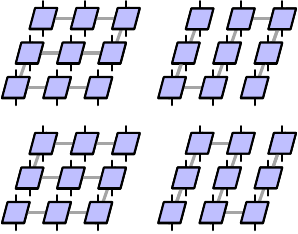}
        \put(0, 75){(a)}
        \put(50, 75){(b)}
        \put(0, 30){(c)}
        \put(50, 30){(d)}
    \end{overpic}
    \caption{%
        Various string configurations in a SnakeSBS.
    }
    \label{appfig:SnakeSBS}
\end{figure}

A SnakeSBS is a string-bond state with multiple MPS strings that weave through the lattice in various snake-like patterns.
\Cref{appfig:SnakeSBS} shows four snake patterns where MPS strings alternate between horizontal and vertical paths.
These patterns are apt at capturing long-range correlations within the lattice.
In our numerical experiments, we employed both SnakeSBS with two strings [Fig.~\ref{appfig:SnakeSBS}(a) and ~\ref{appfig:SnakeSBS}(b)] as well as four strings [Fig.~\ref{appfig:SnakeSBS}(a) through ~\ref{appfig:SnakeSBS}(d)].

\subsection{Restricted Boltzmann machine states}
\label{app:ansatz-rbmdo}

\begin{figure}
    \centering
    \includegraphics[width=0.8\linewidth]{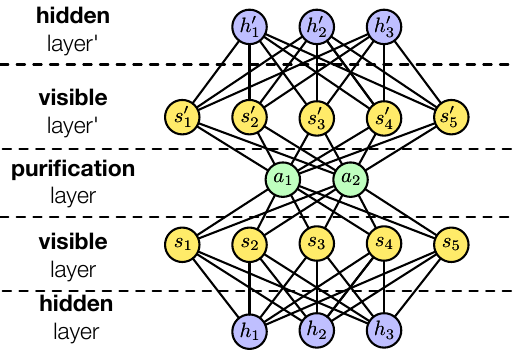}
    \caption{%
        Illustration of a restricted Boltzmann machine (RBM) with a purification layer, depicted as a neural network graph rather than a tensor network diagram, for approximating mixed quantum states.
        The top and bottom \emph{hidden} layers contain neurons that encode correlations between the \emph{visible} physical spins, represented by the middle and bottom layers.
        The \emph{purification} layer in green introduces ancillary degrees of freedom, crucial for constructing a positive semidefinite mixed state via partial tracing.
        The inter-layer connections denote the neural network weights, optimized during training.
    }
    \label{appfig:RBMDO}
\end{figure}

A restricted Boltzmann machine density operator (RBMDO) is obtained from the partial trace over the ancillary degrees of freedom in a restricted Boltzmann machine (RBM).
As depicted in \cref{appfig:RBMDO}, the RBM consists of three layers: the visible layer, which embodies the physical degrees of freedom; the hidden layer, which encodes correlations among physical spins; and the auxiliary layer, which comprises additional degrees of freedom subject to tracing.
The RBM's capacity to represent complex correlations can be tuned by varying the densities of the hidden ($\alpha=N_h/N$) and ancillary ($\beta=N_a/N$) layers relative to the visible layer.
Enhancing these densities enables the RBM to model more intricate correlations at the expense of increased computational requirements.

The RBM purification can be mathematically described by a state that is a product of exponentials of local fields and interactions:
\begin{equation}
    \begin{split}
        \Psi_\text{RBM}(\bm{s},\bm{a}) = &\exp \left(\sum_{j} b_j^s s_j \right)\exp \left(\sum_{k} b_k^a a_j \right)\\
                                         &\times \prod_{i} X_i(\bm{s})\prod_{k} Y_k(\bm{s}, a_k),
    \end{split}
    \label{eq:RBM}
\end{equation}
where
\begin{align}
    X_i(\bm{s})      &= \cosh\left(b_i^{h} + \sum_{j} W_{ij}^h s_j \right),\\
    Y_k(\bm{s}, a_k) &= \exp\left(\sum_{j} W_{kj}^a a_k s_j \right).
\end{align}
In these expressions, $\bm{s}$ and $\bm{a}$ represent the states of the visible and ancillary neurons, which can take values of $\{-1,+1\}$, respectively.
Compared to the main text, here we included the local bias terms, $b_j^s$ and $b_k^a$.
The factors $X_i(\bm{s})$ and $Y_k(\bm{s}, a_k)$ encapsulate the contributions from the hidden and auxiliary layers.
By summing over the degrees of freedom of the hidden layer, and tracing over the auxiliary ones, we obtain a mixed state that is positive semidefinite by construction.
A notable feature of the RBM is that both the sum over hidden neurons and the partial trace over the ancillary degrees of freedom can be performed analytically.

It was shown in Ref.~\cite{Glasser2018NeuralNetwork} that $X_i(\mathbf{s})$ is equivalent to an SBS with diagonal matrices of bond dimension 2. Here we present the derivation.
The key property of the hyperbolic cosine function is:
\begin{equation}
    \cosh(x) = \frac{e^{x} + e^{-x}}{2}.
\end{equation}
We can write $X_i(\bm{s})$ as:
\begin{equation}
    X_i(\bm{s}) = \cosh\left( b_i^h + \sum_j W_{ij}^h s_j \right) = \frac{1}{2} \left( e^{\theta_i(\bm{s})} + e^{-\theta_i(\bm{s})} \right),
\end{equation}
where $\theta_i(\bm{s}) = b_i^h + \sum_j W_{ij}^h s_j$. This expression can be represented as an MPS by introducing auxiliary indices that capture the exponential terms. Specifically, we define local tensors at each site $j$ as:
\begin{equation}
    A^{s_j}_{i,j} = \begin{pmatrix}
        e^{W_{ij}^h s_j+b_i/N } & 0 \\
        0 & e^{-W_{ij}^h s_j-b_i/N}
    \end{pmatrix},
\end{equation}
Then, $X_i(\bm{s})$ can be expressed as an MPS by
\begin{equation}
    X_i(\bm{s}) = \tr\prod_j A^{s_j}_{i,j}
\end{equation}

As shown in the Fig.~\ref{fig:ansatz-rbm}, the tensor network diagram for each of the factor $Y_k(\bm{s}_p,a_p)$ can be represented using the copy tensor [see Fig.~\ref{fig:ansatz-graphical}] as
\begin{equation}
    Y_k(\bm{s}_p,a_k)=
    \vcenter{\hbox{\includegraphics[]{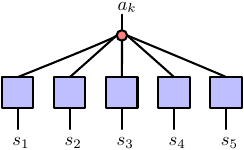}}}\label{appfig:RBM_LVM}.
\end{equation}
This configuration resembles a latent variable model, frequently used in machine learning, where each system spin is independent and only interacts with others via the ancilla.

A RBM accepts only binary inputs $\{-1,+1\}$, which are well suited to the natural states of spin-$1/2$ systems.
Extending this approach to systems with higher spin is less straightforward.
Following the methodology in Ref.~\cite{Glasser2018NeuralNetwork}, we can extend the RBM Ansatz to accommodate larger local dimensions.
This is achieved by directly parameterizing the tensor network diagram [\cref{appfig:RBM_LVM}] with a generalized tensors $T_{a_k,s_j}$, where $a_k$ and $s_j$ serve as tensor indices rather than binary values.
In this extended RBM framework, enlarging the ancilla's dimension effectively corresponds to incorporating additional ancillae.

In a local or short-range RBM, each hidden and auxiliary neuron only connects to a fixed number of visible spins~\cite{Glasser2018NeuralNetwork}.
The graphical representations given by \cref{appfig:RBM_LVM,appfig:EPS_puri} show that local RBM are a subset of local EPS.
Although local EPS are more versatile than local RBM, they require exponentially more parameters relative to the size of each plaquette, increasing computational complexity.
Local RBM adhere to the area law for both entanglement entropy in their purification and mutual information in the corresponding mixed states.
Conversely, fully-connected RBM disregard the spatial lattice structure and the inherent locality of quantum states, leading to a volume law for entanglement entropy in the purification~\cite{Deng2017Quantum}, as well as for mutual information in the mixed states they represent.
This behavior contrasts with the typical area law scaling of entanglement entropy and mutual information found in the ground and thermal states of local gapped Hamiltonians.
Hence it can be expected that certain RBM states cannot be represented efficiently using local tensor networks such as MPS and PEPS.

\section{Numerical Implementation}
\label{app:numerical-implementation}

Here, we provide specifics of our numerical implementation.
We discuss techniques to speed up the calculations of the gradients (\cref{app:gradient}), the sampling process (\cref{app:sampler}) and the choice of the optimizer (\cref{app:optim}).
All simulations were performed using GPU-specific kernels on NVIDIA P100/V100/A100 graphic cards.

\subsection{Gradients}
\label{app:gradient}

In the numerical implementation of our variational Monte Carlo algorithm, estimating the gradients of $F_R(\rho)$ requires the computation of $\rho_{\bm{ss}}$ as well as the gradient with respect to the parameters, $\partial_{\theta} \rho_{\bm{ss'}}$.
First, since tensor network Ans\"atze require mostly tensor contraction operations (\texttt{jax.numpy.einsum}), they can be automatically batched with \texttt{jax.numpy.vmap}.
Since modern GPUs are highly optimized for performing tensor contractions, their evaluations is highly parallelizable on GPUs, with efficiency comparable to that of neural networks.
The derivatives are computed with automatic differentiation with existing deep learning libraries~\cite{Jax}.
When the model or sample size is large, we compute the gradients in chunks to reduce the memory footprint.
The implementation is adapted from existing chunking utilities of Netket 3~\cite{Vicentini2022NetKet}.
When the model requires a series of sequential contractions, we have found it beneficial to also implement the gradient checkpointing~\cite{Chen2016Training} technique to reduce the number of cached intermediate results in the propagations.
This allowed us to scale up the simulations to larger bond dimensions.

RBM states with real parameters can only represent quantum states where all amplitudes are positive.
Even for real-valued states, in order to represent the sign structure---which arises in any non-stoquastic Hamiltonian---complex parameters must be used.
Equivalently, one may split the complex parameters into two sets of real parameters: one set representing the amplitudes, and the other set representing the phase.
Handling complex parameters can introduce additional computational complexity.
Specifically, when computing the gradients and applying the Gram matrix, one needs to calculate four real vector-Jacobian products (\texttt{jax.vjp} in Jax~\cite{Jax}), due to the real and imaginary parts of the complex parameters.
In contrast, all of our tensor network Ans\"atze do not have this problem.
Since they naturally handle real-valued states, there is no need to introduce complex variables for the sign structure.
This simplifies the computation and results in a three-fold speed increase in the computation of the gradients for the tensor network Ans\"atze ($\mathbb{R}\to \mathbb{R}$) compared to the neural network Ans\"atze ($\mathbb{C}\to \mathbb{C}$).

\begin{figure*}[th]
    \centering
	\includegraphics[width=0.8\linewidth]{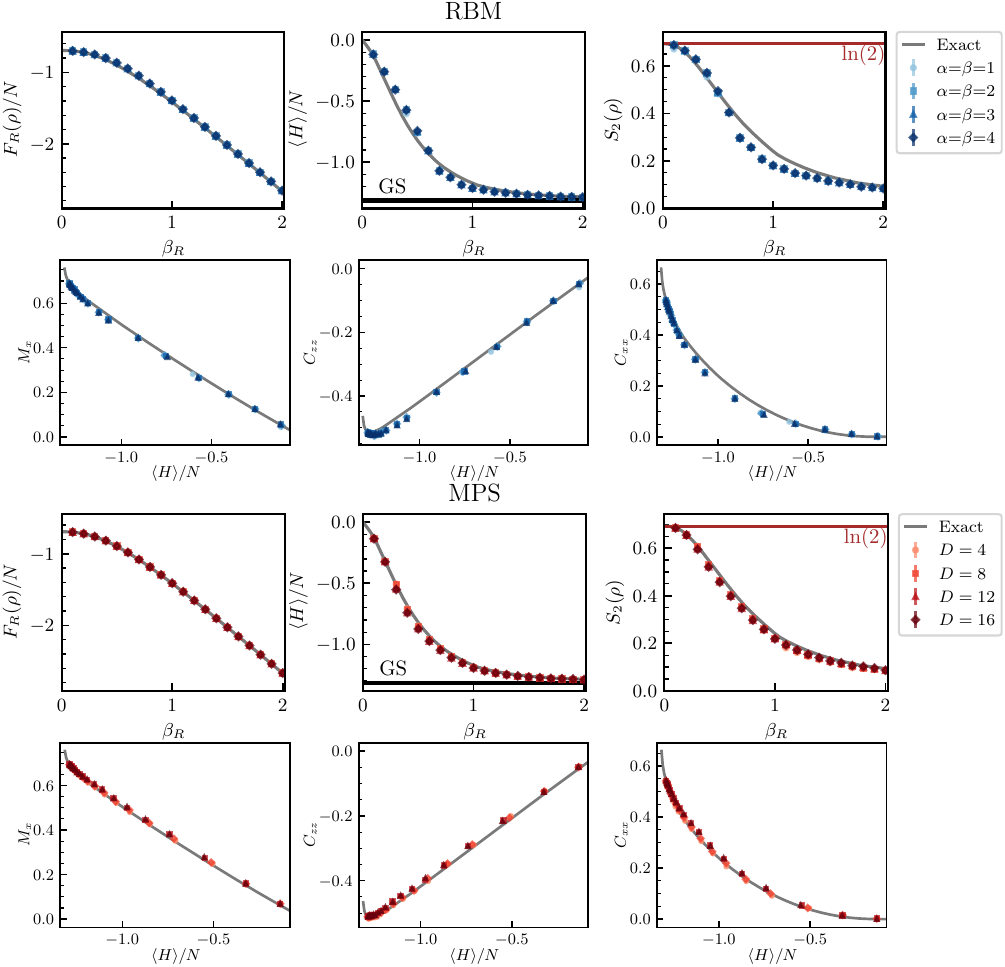}
    \vspace*{-2mm}
    \caption{%
        Extended analysis of the non-integrable Ising model with parameters $J=1.0$, $h_x=1.05$, and $h_z=0.5$ for a system size of $N=16$ spins, plotted as a function of the inverse temperature $\beta_R$.
        The top panel illustrates the performance of the restricted Boltzmann machine (RBM) Ans\"atze for different values of $\alpha=\beta$, showing a saturation in the variational free energy per site, the energy per site, and the R\'enyi entropy per site as $\alpha=\beta$ increases.
        The bottom panels display the results for the MPDO Ans\"atze with different bond dimensions $D$.
    }
    \label{appfig:1d-results-nonintegrable-3}
\end{figure*}

\subsection{Samplers}
\label{app:sampler}
We need to sample from two distinct probability distributions.
The first one is a distribution on the diagonal elements of the density matrix, whose probability density is proportional to the diagonal elements $\rho_{\bm{ss}}$ of the density matrix.
The second one is a distribution over the off-diagonal elements of the density matrix, whose probability density is proportional to $|\rho_{\bm{ss'}}|^2$.
We use the Metropolis-Hastings algorithm to generates a Markov chain of configurations with the desired stationary distribution.
For the diagonal elements, we adapt the standard update rule used in VMC methods for sampling pure state configurations.
Common update rules for the diagonal elements include local spin flips and local exchanges of pairs~\cite{Vicentini2022NetKet}.
The update rules for off-diagonal elements need to account for the possible quasi-diagonal form of the density matrix near infinite temperature.
We find that the following approach maintains a reasonable acceptance ratio near the $50\%$:
\begin{enumerate}
    \item With a 50\% probability, we flip one index either in the $\bm{s}$ or $\bm{s'}$ configuration, which corresponds to an off-diagonal update.
    \item With a 50\% probability, we flip the same index for the $\bm{s}$ and $\bm{s'}$, which corresponds to a diagonal flip.
\end{enumerate}

\subsection{Optimizer}
\label{app:optim}

Once the gradients with respect to the parameters [\cref{eq:gradient}] are obtained, first-order optimizers can be applied to minimize $F_R(\rho)$.
We use the Adam optimizer~\cite{Kingma2015Adam}, typically with a learning rate of 0.01.
Sometimes, the gradient descent method [cf.\ \cref{eq:GD}] can lead to significant changes in the variational state with minor changes in parameters $\theta$.
In variational Monte Carlo, this issue can be addressed by the stochastic reconfiguration method, which takes into account the natural metric in the physical state space.
We use a similar stochastic reconfiguration method for mixed states as described in \cref{app:FlowEqorSRmixed}.
To enhance the numerical stability of the optimization process, we add a small diagonal shift of $\epsilon=0.01 \text{--}0.1$ to the Gram matrix.
This shift helps to prevent numerical instabilities that can arise from the inversion of the Gram matrix during the optimization process.

Our optimization process is designed to ensure the convergence of the R\'enyi free energy and the stability of the variational parameters.
The process typically runs between $1000$ and $10\,000$ iterations.
We define energy convergence as the point at which the change in energy between iterations falls below a certain threshold.
The optimization process stops when either the maximum number of iterations is reached or the energy convergence is achieved.
The optimization process consists of several stages:
\begin{itemize}
    \item \emph{Warm-up stage:} During the initial phase of optimization, we start from a random initial state.
          The step size is gradually increased, allowing the optimization process to explore a larger portion of the parameter space.
    \item \emph{Convergence stage:} As optimization progresses, we reduce the step size.
          This reduction in step size helps to ensure the convergence of the optimization process by allowing it to fine-tune the variational parameters.
    \item \emph{Initialization with smaller models:} We found that initializing larger models with optimized smaller models consistently improves the quality and stability of the optimization process.
          This strategy uses the solutions of smaller models to provide a good starting point for the optimization of larger models, which is reminiscent of how bond dimensions are enlarged in DMRG.
\end{itemize}

Given the computational demands of the optimization process, the free energy is not estimated at each step.
Instead, once the optimization converges, the energy, a set of relevant physics observables, and the R\'enyi entropy of the variational state are estimated.
This estimation is performed using a large number of samples, on the order of $2^{16}$, to ensure statistical accuracy.

\section{Additional numerical results}
\label{app:additional-numerical-results}

We present additional numerical results with further comparisons of the variational Ans\"atze across different models and parameters for both the one-dimensional and two-dimensional problems.

\subsection{Results on the 1D Ising models}
\label{app:1D-results}

\cref{appfig:1d-results-nonintegrable-3} provides an extended analysis of the non-integrable Ising model considered in the main text.
The upper panels illustrate the performance of the RBM Ansatz with different neural densities $\alpha$ and $\beta$, while the lower panels display the results for the MPDO Ans\"atze with varying bond dimensions $D$.
In contrast to the RBM Ansatz, the MPDO Ans\"atze demonstrates a systematic improvement in capturing the thermal state properties as the bond dimension increases, as evidenced by the convergence toward the exact diagonalization results.

Additionally, we have also run benchmarks for the integrable one-dimensional transverse field Ising model with parameters $J=1,\ h_z=0,\ h_x=1$ (which is critical at zero temperature) and observed similar trends for both the RBM and MPS Ans\"atze in the variational free energy, energy, and R\'enyi entropy per site.

\subsection{Results on the 2D transverse field Ising model}
\label{app:2D-results-nonintegrable}

To compare the performance of different Ans\"atze, we consider the two-dimensional quantum transverse field Ising model of Fig.~\ref{fig:2d-results-a}.
The R\'enyi free energies at a particularly challenging $\beta_R$ are summarized in \cref{table:2D_results}. The comparison is made against exact diagonalization results for the same system size $N=16$ and the same field strength $h_x=3.0$.

\begin{table}[hbt]
    \centering
    \resizebox{0.45\textwidth}{!}{
        \begin{tabular}{@{}lllllll@{}}
            \toprule
                              & $\alpha$=$\beta$=1 & $\alpha$=$\beta$=2 & $\alpha$=$\beta$=3 & $\alpha$=$\beta$=4 &   & \\
            RBM               & -20.433            & -20.428            & -20.437            & -20.432            &     \\
            \midrule
                              & $D$=2              & $D$=3              &                    &                    &     \\
            PEPS              & -20.420            & -20.541            &                    &                    &     \\
            \midrule
                              & $D$=4              & $D$=8              & $D$=16             & $D$=32             &     \\
            MPS               & -20.119            & -20.405            & -20.715            & -20.813            &     \\
            \midrule
                              & $D$=4              & $D$=8              & $D$=12             & $D$=16             &     \\
            SnakeSBS($n_s$=2) & -20.346            & -20.420            & -20.468            & -20.510            &     \\
            \midrule
                              & $D$=4              & $D$=8              & $D$=12             & $D$=16             &     \\
            SnakeSBS($n_s$=4) & -20.672            & -20.765            & -20.795            & -20.799            &     \\
            \midrule
                              & $w,\chi$=2,16      &                    &                    &                    &     \\
            EPS               & -20.380            &                    &                    &                    &     \\

            \bottomrule
        \end{tabular}}
    \caption{%
        R\'enyi free energy values for different Ans\"atze.
        The minimal R\'enyi free energy, determined by exact diagonalization, is approximately $F_{\min}^{ED} \simeq -20.91$.
    }
    \label{table:2D_results}
\end{table}

\begin{figure*}[hp]
	\includegraphics[width=0.8\linewidth]{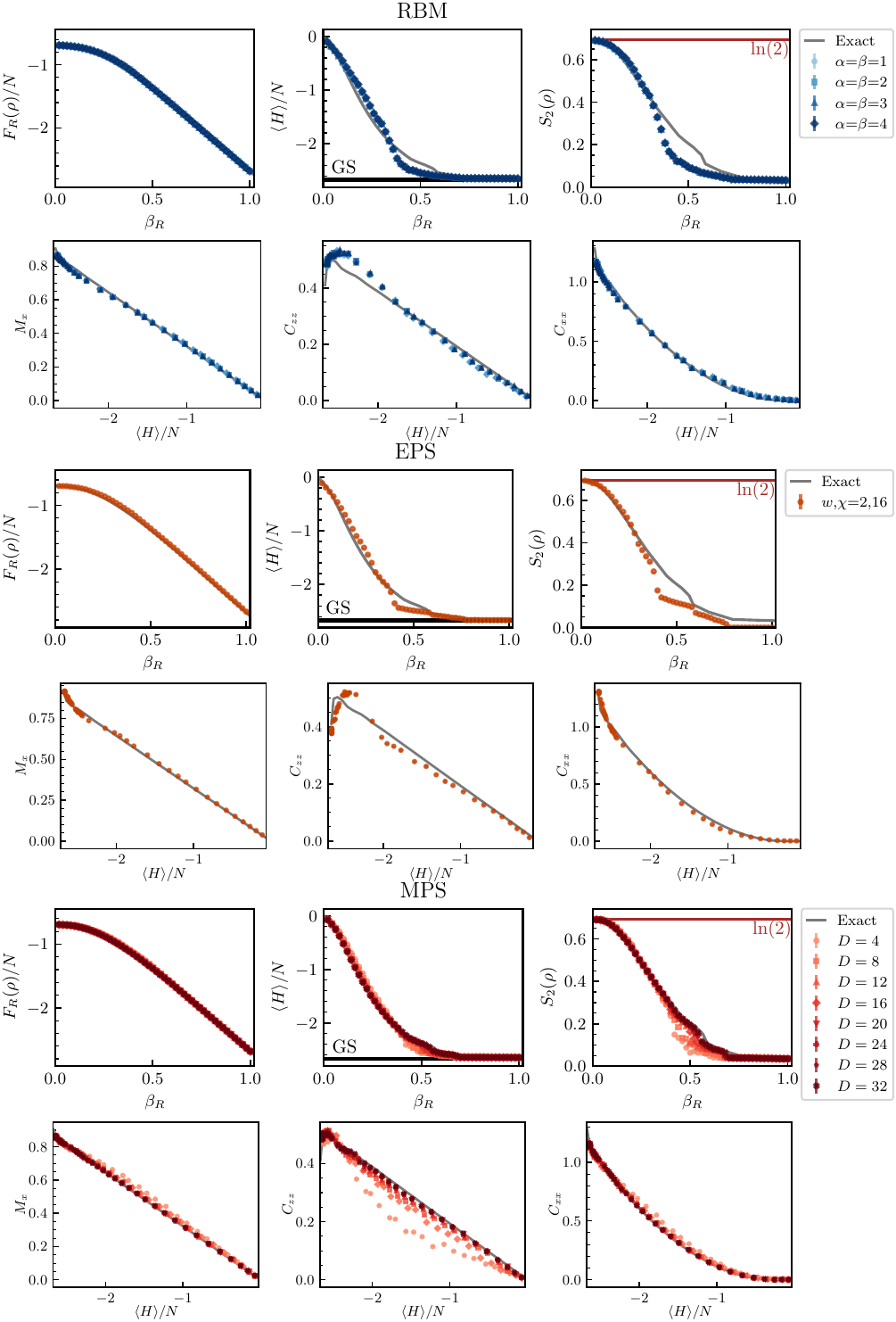}
    \caption{The performance of different Ans\"atze (I: RBM, MPS, EPS)  for approximating the thermal states of the two-dimensional transverse field Ising model with parameters $J=-1.0$, $h_z=0$, and $h_x=2.5$.}
    \label{appfig:2d-results}
\end{figure*}

\begin{figure*}[hp]
    \centering
    \includegraphics[width=0.8\linewidth]{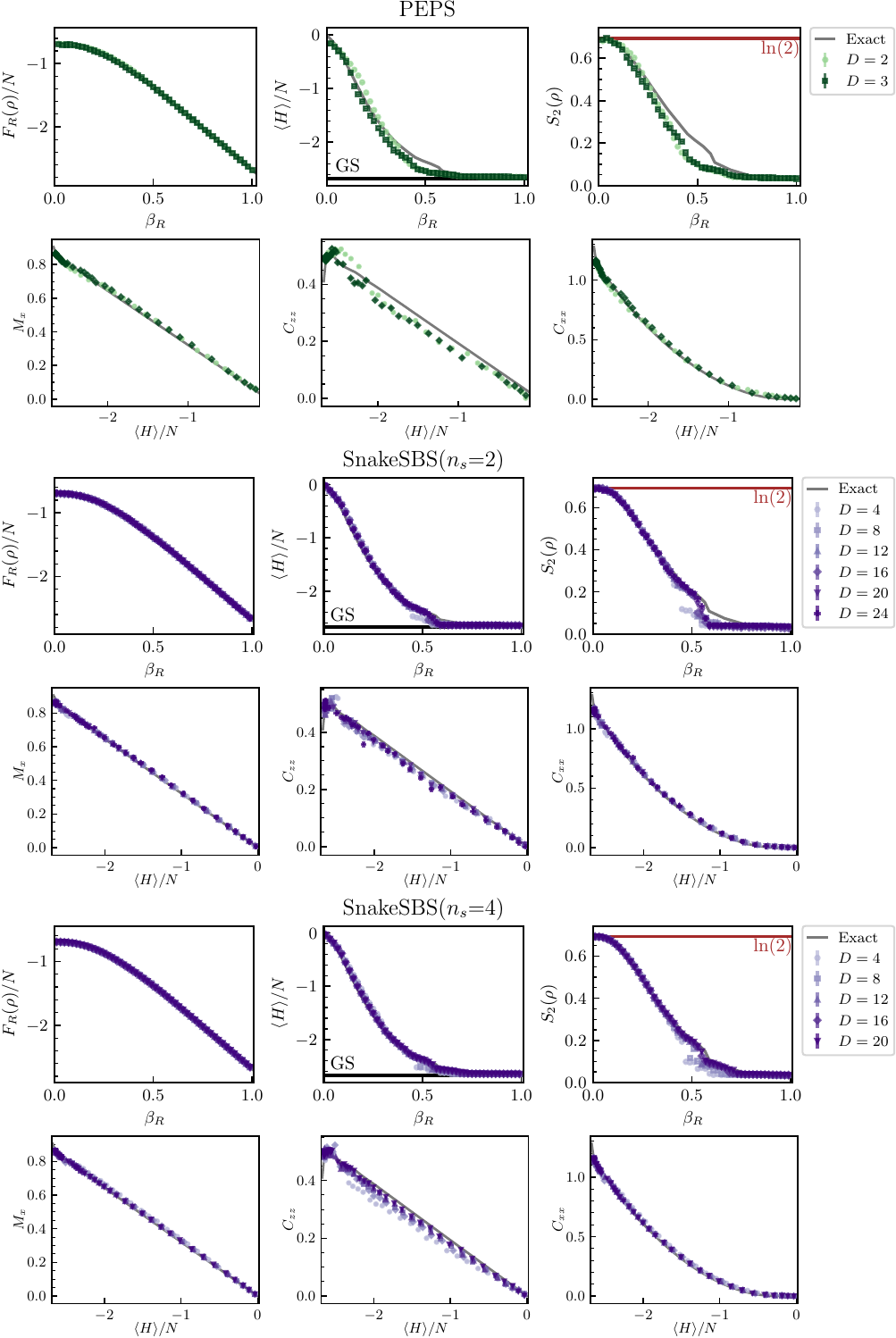}
    \caption{The performance of different Ans\"atze (II: PEPS, SnakeSBS($n_s$=2), SnakeSBS($n_s$=4)) for approximating the thermal states of the two-dimensional transverse field Ising model with parameters $J=-1.0$, $h_z=0$, and $h_x=2.5$.}
    \label{appfig:2d-results-snake}
\end{figure*}

Further away from the critical field, we display the other quantities involved in the cost function over a range of values of $\beta_R$ in \cref{appfig:2d-results} and \cref{appfig:2d-results-snake}. The comparison is made against exact diagonalization results for the same system size $N=16$ and the same field strength $h_x=2.5$ as Fig.~\ref{fig:2d-results-b}.

%

\end{document}